\documentclass[aps,prx,longbibliography,nofootinbib,superscriptaddress,amsmath,amssymb,twocolumn,floatfix]{revtex4-1}
\usepackage{bm}
\usepackage{soul}
\usepackage{bbold}
\usepackage{datetime}
\usepackage[normalem]{ulem}
\usepackage{graphicx,color}
\usepackage{graphicx,scrtime}
\usepackage{tcolorbox}

\newcommand{\Sig}{\widetilde \Sigma}

\newcommand{\kk}{{\bf k}}

\newcommand{\uu}{{\bf u}}
\newcommand{\vv}{{\bf v}}

\newcommand{\rr}{{\bf r}}

\newcommand{\FF}{{\cal F}}
\newcommand{\NN}{{\cal N}}

\newcommand{\VV}{{\cal V}}

\newcommand{\be}{\begin{equation}}
\newcommand{\ee}{\end{equation}}
\newcommand{\ba}{\begin{eqnarray}}
\newcommand{\ea}{\end{eqnarray}}
\newcommand{\bse}{\begin{subequations}}
\newcommand{\ese}{\end{subequations}}
\newcommand{\beq}{\begin{eqnarray}}
\newcommand{\eeq}{\end{eqnarray}}

\newcommand{\ts}{\textstyle}

\begin{document}
\title{Instantaneous normal modes in liquids: a heterogeneous-elastic-medium approach}
\author{Stefano Mossa}
\affiliation{Univ. Grenoble Alpes, CEA, IRIG-MEM-LSim, 38054 Grenoble, France}
\author{Taras Bryk}
\affiliation{Institute for Condensed Matter Physics,National Academy of Sciences of Ukraine, UA-79011 Lviv, Ukraine} 
\affiliation{Institute of Applied Mathematics and Fundamental Sciences, Lviv National Polytechnic University, UA-79013 Lviv, Ukraine} 
\author{Giancarlo Ruocco}
\affiliation{Center for Life Nano Science @Sapienza, Istituto Italiano di Tecnologia, 295 Viale Regina Elena, I-00161, Roma, Italy}
\affiliation{Dipartimento di Fisica, Universita' di Roma ``La Sapienza'', P'le Aldo Moro 5, I-00185, Roma, Italy}
\author{Walter Schirmacher}
\affiliation{Center for Life Nano Science @Sapienza, Istituto Italiano di Tecnologia, 295 Viale Regina Elena, I-00161, Roma, Italy}
\affiliation{Institut f\"ur Physik, Staudinger Weg 7, Universit\"at Mainz, D-55099 Mainz, Germany}
\begin{abstract}
 Since decades, the concept of vibrational density of states in glasses has been mirrored in liquids by the instantaneous-normal-mode spectrum. While in glasses instantaneous configurations correspond to minima of the potential-energy hypersurface and all eigenvalues of the associated Hessian matrix are therefore positive, in liquids this is no longer true, and modes corresponding to both positive and negative eigenvalues exist. The instantaneous-normal-mode spectrum has been numerically investigated in the past, and it has been demonstrated to bring important information on the liquid dynamics and transport properties. A systematic deeper theoretical understanding is now needed. Heterogeneous-elasticity theory has proven to be particularly successful in explaining many details of the low-frequency excitations in glasses, ranging from the thoroughly studied boson peak, down to the more elusive non-phononic excitations observed in numerical simulations at the lowest frequencies. Here we present an extension of heterogeneous-elasticity theory to the liquid state, and show that the outcome of the theory agrees well to the results of extensive molecular-dynamics simulations of a model liquid at different temperatures. We find that the spectrum of eigenvalues $\rho(\lambda)$ has a sharp maximum close to (but not at) $\lambda=0$, and decreases monotonically with $|\lambda|$ on both its stable and unstable side. We show that the spectral shape strongly depends on temperature, being symmetric at high temperatures and becoming rather asymmetric at low temperatures, close to the dynamical critical temperature. Most importantly, we demonstrate that the theory naturally reproduces a surprising phenomenon, a zero-energy spectral singularity with a cusp-like character developing in the vibrational spectra upon cooling. This feature, known from  a few previous numerical studies, has been generally overlooked in the past due to a misleading representation of the data. We provide a thorough analysis of this issue, based on both very accurate predictions of our theory, and computational studies of model liquid systems with extended size.
\end{abstract}
\maketitle
\section{Introduction}
\label{sect:introduction}
Clarifying the properties of the potential-energy hypersurface (or potential-energy landscape, PEL) and its connection with the (slow) dynamics and transport properties in supercooled liquids is a hot topic since decades (see, among many others,~\cite{keyes89,stratt95,bembenek95,keyes97,sciortino97,sastry01,taraskin02a,debenedetti2001supercooled,cavagna01,broderix00,Angelani00,angelani00a,angelani02,grigera02,fabricius02,ciliberti03,sampoli03,angelani03,berthier03,doye02,doye03,doye03a,doliwa03,grigera06,clapa12,berthier19,zhang19}).  The  idea traces back to Goldstein~\cite{goldstein69} who described the PEL as an ensemble of basins of attraction, each one associated with a local minimum (the inherent structure, IS). The dynamics was thus described as the interplay of (quasi-)harmonic motion confined around minima, and the jumps (more or less rare, depending on temperature, $T$) between the basins of attraction associated with different minima. 

This picture was further developed by Stillinger and Weber~\cite{stillinger84,stillinger95}. They defined the basin of attraction of a given IS (i.~e., the set of points connected to the minimum through a steepest-descent path) in such a way that the basins of attraction generate a complete description of the configurational space. Next, they wrote the partition function as $Z=\sum_{\{b\}} Z_{b}$, where $\{b\}$ indicates the set of all the basins, and $Z_{b}$ is the partition function associated to the dynamics within the basin $b$. As a consequence, the free energy of a liquid could be expressed~\cite{stillinger84,stillinger95} as the sum of the free energy of a "typical" basin, $\FF_{b}(T)$, and a term accounting for the multiplicity of basins accessible by the system, $\NN(T) = \exp(\Sigma(T)/k_B)$, with $\Sigma(T)$ the configurational entropy. According to these ideas, the thermodynamics of the liquids could therefore be described in terms of the {\it minima} of the PEL.

A further step forward was proposed by Keyes~\cite{keyes89,keyes97}, who used the instantaneous normal modes (INM) to predict not only the very short time dynamics within a basin, but also the long time diffusion processes of crossing the basins' borders. In the INM approach, the local curvature of the PEL is calculated along equilibrated molecular-dynamics (MD) trajectories. The idea is to recover the global shape of the landscape from a collection of local curvatures, specifically those that are sampled during the equilibrium dynamics. For systems evolving close to a local minimum, the local curvatures will be all positive. In contrast, points near the borders separating basins will have at least one negative curvature, most likely more than one. Intuitively, one can expect that the larger the amount of negative curvatures present, the easier the inter-basins jumps, thus inducing a larger diffusion coefficient, $D$. Indeed, although the exact functional form is system dependent, the INM approach suggests that $D$ is a monotonic function of the fraction of unstable modes.

The connection between the properties of the INM and the diffusion coefficient in supercooled liquids immediately points towards the physics of the structural arrest, and the theories for the glass transition(see, among other reviews,~\cite{berthier2011theoretical}). Among them, we mention the mode-coupling theory (MCT)~\cite{gotze2009complex}, a mean-field approach which predicts the existence of a finite dynamical critical temperatures, $T_{MCT}$, where $D=0$. The study of the INM reveals~\cite{angelani00a,broderix00} that near $T_{MCT}$ the fraction of unstable modes vanishes. One can therefore conclude that at $T_{MCT}$ the system ceases to evolve near the borders of the basins of attraction, starting to spend most of the time close to minima.

An analysis made by Kurchan and Laloux~\cite{kurchan96} (KL) may help to elucidate the above process. KL focused on the properties of the borders of the basins of attraction in high-dimensional configuration spaces. They concluded that in an $N$-dimensional space a basin has a volume at least of the order $\ell^N$, where $\ell$ is the typical distance between configurations pertaining to the same basin. This has as a consequence that in the thermodynamic limit, $N \rightarrow \infty$, most of the volume associated to the basins is located near their borders, away from the minima. KL then hypothesized that, at high $T$, the system populates regions of borders belonging to many basins. Near $T_{MCT}$, in contrast, it is mainly situated in regions of borders of a few basins, which obviously encompass smaller volumes compared to the former case. 

Interestingly, the KL analysis points towards a hierarchical organization of the configuration space. Indeed, let us assume that we were able to study the potential energy $U$ restricted to the borders, $\partial b$, of the basins at low $T$. This restricted PEL is a $(N-1)$-dimensional manifold where all the stationary points have their order decreased by one. As a consequence, saddles of order one of the original PEL become local minima of the restricted PEL. Similarly, at higher $T$ the trajectories will be localized close to the $(N-2)$-dimensional borders of $(N-1)$-dimensional basins of attractions. One can therefore iterate the procedure $k$ times until reaching the relevant $(N-k)$-dimensional manifold in the vicinity of which equilibrium trajectories are ``almost always'' within some small distance. This analysis overall demonstrates the importance of saddle points (of increasing order on increasing $T$) for the description of the liquid dynamics, and, specifically, the importance of saddles of order one for the dynamics close to the arrest temperature.

Following this line of thought, Angelani {\it et al.}~\cite{angelani00a} and  Broderix {\it et al.}~\cite{broderix00} studied the role played by saddles in the supercooled liquid dynamics using a trick to find saddles of order one, i.e., by searching for the stationary points of the function $\Phi= \vert \vec \nabla \VV \vert^2$, with $\VV$ the system potential energy. They found  that the order of the saddles is a decreasing function of the temperature and it vanishes at $T_{MCT}$. This numerical result pinpoints the physical meaning of $T_{MCT}$: it is the temperature marking the cross-over from a dynamics taking place near the basins' borders, to a dynamics mainly localized within the basins. 

Despite a large body of literature focusing on the INMs, and on the pivotal role they play in structural arrest, a robust theory describing the main features of the density of the instantaneous normal modes (including a surprising zero-energy spectral singularity at $T_{MCT}$, see below) is still lacking. Here, following our recent work~\cite{schirm22}, we present a theoretical description of the INM spectra in liquids, accompanied by a extensive numerical simulation of a model system. In our theory, a generalization of the heterogeneous-elastic theory (HET) solved in the self-Consistent Born approximation (SCBA)~\cite{Schirmacher2006,Schirmacher2007,Marruzzo13,schirm14}, the instantaneous-normal-mode spectrum of the liquid is described as that of an elastic medium, where the local shear moduli exhibit strong spatial fluctuations, including a large number of negative values. Realizing that these fluctuations are merely snapshots of thermal fluctuations~\cite{mizuno2020sound,mizuno2019impact}, we assume that they obey Gaussian statistics, with a variance proportional to the temperature, thus introducing the temperature dependence of the INM in a very natural way. The outcome of the theory is shown to favorably compare with the results of the extensive MD simulations.
\section{Theory}
\label{sect:theory}
We consider a system formed by $N$ particles with a potential energy $\VV(\{\rr_i(t)\}_{i=1..N})$, which depends on the instantaneous positions $\rr_i(t)$
of the particles at time $t$. We assume that it can be expressed in terms of pairwise potentials $\phi(r_{ij}(t))$ as,
\be
\VV(\{\rr_i(t)\}_{i=1..N})=\frac{1}{2}\sum_{i\neq j}\phi(r_{ij}(t)),
\label{PotentialPairwise}
\ee
with $r_{ij}(t)=\vert \rr_i(t)-\rr_j(t) \vert$. In the INM procedure one considers virtual small displacements, $\uu_i(t)$, around the instantaneous particle positions, $\rr_i(t)$, obtained in a MD simulation. The second-order coefficients of a Taylor expansion of $\VV$ form the Hessian matrix,
\be
\label{in1}
H_{ij}^{\alpha\beta}(t)=\frac{\partial^2}{\partial r_i^{\alpha}\partial r_j^{\beta}}
\VV(\{\rr_1(t)\dots\rr_N(t)\}),
\ee
($\alpha, \beta=1,\ldots, 3$), which can be diagonalized to obtain the eigenvalues $\lambda_p$, with $p=1,\dots, 3(N-1)$~\footnote{3 eigenvalues are trivially zero, due to  the conditions
\mbox{$\sum_i H_{ij}^{\alpha\beta}=0$.}}, together with the associated spectrum, $\rho(\lambda_p)$.

If the system is close enough to a minimum, all the curvatures are positive and so are all the $\lambda_p$. These are related to the square of the vibrational frequencies, $\lambda_p = \omega_p^2$, and one can consider the Hessian as the counterpart of the dynamical matrix of a solid. In this regime one can therefore identify the spectrum with the density of vibrational states (DOS) of the system,
\be
\label{in2}
g(\omega)=2\, \omega \rho(\lambda(\omega))=\frac{\ts 1}{\ts 3N-3} \sum_{p=1}^{3N-3} \delta(\omega - \omega_p)\, .
\ee
If, in contrast, the system is away from a minimum, one or more negative curvatures exist, and the non-trivial eigenvalues of the Hessian are both positive and negative. In this case the identification of $\lambda_p$ with the square of a "frequency" is obviously no longer possible. For the unstable part of the spectrum ($\lambda_p \le 0$) it has become common practice to define the positive quantity $\tilde\omega = -i \sqrt{\lambda}$, together with the corresponding ``DOS'' as,
\be
\label{in3}
g(\tilde\omega)\equiv 2\, \tilde\omega\rho(\lambda(\tilde\omega)).
\ee
This function is usually plotted on the negative $\omega$-axis, and the stable $g(\omega)$, according to Eq.~(\ref{in2}), on the positive $\omega$-axis. 

This representation, although commonly used as an alternative to directly looking at the $\rho(\lambda)$, is misleading as unfortunately hides many crucial details of the INM spectrum. Indeed, the former is obtained from the latter by multiplication by $|\omega|$, a procedure, which strongly suppresses all low-frequency details of the spectrum $\rho(\lambda)$ \cite{schirm22}. Based on this observation, Sastry {\it et al.}~\cite{sastry01}, and Taraskin and Elliott ~\cite{taraskin02a} showed that the INM spectrum of liquids, when plotted directly as $\rho(\lambda)$ exhibits a characteristic cusp-like maximum at $\lambda=0$, which cannot be seen in an analysis based on Eqs.~(\ref{in2}) and~(\ref{in3}).

In order to obtain a theoretical description of the spectrum $\rho(\lambda)$, we start from the equations of motion of linear elasticity with a spatially fluctuating shear modulus, $G(\rr)$, similar to HET in glasses. HET has been rather successful in explaining several low-frequeny anomalies in the vibrational spectrum of glasses~\cite{Schirmacher2006,Schirmacher2007,Marruzzo13,schirm14,pan21}, including the so-called Boson peak, an enhancement of the DOS at finite frequencies over the Debye prediction $g(\omega)\propto \omega^2$. We now aim to explore the application of this theory in the unstable regime, and compare the results with the INM spectrum of a simulated soft-sphere liquid (see below). 

We consider the equations of motion for the continuum version of the virtual displacements, $\uu_i(t)\rightarrow \uu(\rr,t)$. The time derivative of this quantity, $\vv(\rr,t)=\dot\uu(\rr,t)$, is the Eulerian instantaneous local velocity field. The HET equation of motion is,
\be
\label{eqmo1}
\rho\ddot\uu(\rr,t)=
\nabla M(\rr)\nabla\cdot\uu(\rr)-\nabla\times G(\rr)\nabla\times\uu(\rr),
\ee
where $M(\rr)=K_0+\frac{4}{3}G(\rr)$ is the longitudinal modulus. Here, $K_0$ is the bulk modulus, which is considered not to exhibit relevant spatial fluctuations. In contrast, the fluctuating shear modulus is written as $G(\rr)=G_0-\Delta(\rr)$, where $G_0$ is the spatial average of $G(\rr)$, and $\Delta(\rr)$ is a local deviation from the average, which we assume to follow a  normal distribution with a standard deviation $\sigma^2$ given by
\be
\label{standard}
\sigma^2=\langle\Delta(\rr)^2\rangle
\ee

In the HET theory for a glass~\cite{Schirmacher2006,Schirmacher2007,Marruzzo13,schirm14,pan21}, the standard deviation $\sigma$ is, of course, temperature {\it independent}. In the present treatment for a liquid, in contrast, one expects a temperature {\it dependent} $\sigma$, as the features of the visited regions of the PEL depend on $T$. By comparing with the simulation data we shall see that $\sigma^2$ is proportional to the temperature, a feature already observed earlier in a simulation of a glass-forming liquid at higher temperatures~\cite{mizuno2019impact}.

To proceed further, we decompose the displacements into the longitudinal ($\nabla\times\uu_L=0$) and transverse \mbox{($\nabla\cdot\uu_T=0$)} components, and transform to frequency space, leading to the stochastic Helmholtz equations,
\be
\label{helm2}
\omega^2\uu_{L,T}(\rr,\omega)=\nabla v_{L,T}^2(\rr)\nabla^2\uu_{L,T}(\rr,\omega).
\ee
Here, $v_L^2(\rr)=M(\rr)/{\rho}$ and $v_T(\rr)^2=G(\rr)/{\rho}$ are the locally fluctuating squared sound velocities, longitudinal (L) and transverse (T), respectively, and $\rho$ is the mass density. Considering Gaussian fluctuations for $\Delta(\rr)$, we can use the replica trick to calculate the average spectrum~\cite{mckane81,john83,Schirmacher2006}.
\subsubsection{The Self-Consistent Born Approximation (SCBA)}
\label{subsubsect:scba}
From a saddle-point approximation of the resulting effective action, we obtain an effective-medium theory where the fluctuating quantity $\Delta(\rr)$ is replaced by a complex frequency-dependent self-energy, $\Sigma(z)$, with $z=\omega^2+i\epsilon=\lambda+i\epsilon$, for $\epsilon\rightarrow 0_+$. $\Sigma(z)$ obeys the self-consistent SCBA Eq.~(\ref{self1})~\cite{Schirmacher2006,Schirmacher2007,Marruzzo13,schirm14}, see below.

\begin{figure}[b]
\centering
\includegraphics[width=0.49\textwidth]{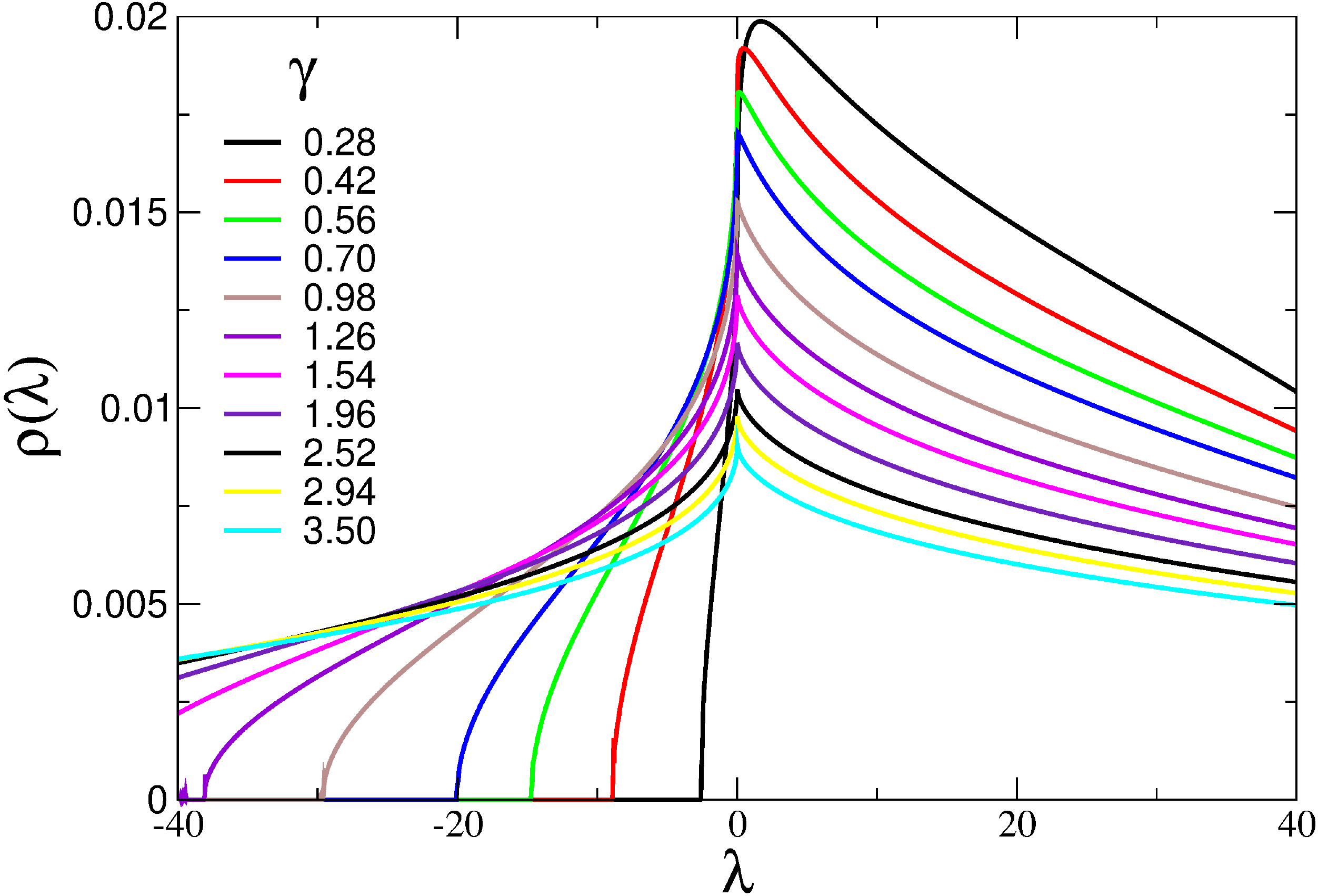}
\caption{INM spectra, $\rho(\lambda)$, of our unstable-elasticity model, calculated in the self-consistent Born approximation (SCBA) as described in the main text. We have considered different values of the disorder parameter $\gamma$, all of them in the liquid side ($\gamma > \gamma_c$=1/4), as indicated in the legend. For better comparison with the numerical data of Fig.~\ref{inm1} we converted the eigenvalues to LJ units according to $\lambda=40\,\tilde\lambda$.}
\label{FigA}
\end{figure}

In three dimensions, the longitudinal and transverse Green's functions for Eqs.~(\ref{helm2}) are,
\be
{\cal G}_{L,T}(k,z)=\frac{1
}{-z+k^2v_{L,T}(z)^2
},
\ee
where the $v_{L,T}(z)$ are the effective complex frequency-dependent sound velocities,
\ba
v^2_{L}(z)&=&\frac{1}{\rho}
\big[K_0+\frac{4}{3}\big(G_0-\Sigma(z)\big)\big]\\
v^2_{T}(z)&=&\frac{1}{\rho}\big[G_0-\Sigma(z)\big)\big].\nonumber
\ea
We choose $G_o$ as the reference shear modulus $G_0$, which is the average shear modulus in the stable glass. $G_0$ might be approximately correspond to $G_\infty$, the liquid high-frequency shear modulus, which appears in the Maxwell relation $\tau=\eta/G_\infty$ between the relaxation time $\tau$ and the shear viscosity $\eta$~\cite{Litov}. The self-consistent HET-SCBA equation for the self energy $\Sigma(z)$ is,
\ba
\label{self1}
\Sigma(z)&=&\sigma^2 V_c
\,{\ts \left(\frac{1}{2\pi}\right)^3} \; \frac{1}{\rho}
\nonumber\\
&\times&
\int_{|\kk|\leq k_\xi} 
d^3\kk \; k^2\bigg(
\frac{2}{3}{\cal G}_{L}(k,z)+{\cal G}_{T}(k,z)
\bigg),
\ea
where $V_c$ is the coarse-graining volume, and the upper cutoff, $k_\xi$, is inversely proportional to the correlation length, $\xi$, of the fluctuations. We now introduce the dimensionless variables, $q$=$k/k_\xi$, $\Sig(z)$=$\Sigma(z)/G_0$, $\tilde K_0$=$K_0/G_0$, $\tilde M(\lambda)$=$\tilde K+\frac{4}{3}[1-\Sig(\lambda)]$, and the (also dimensionless) {\it disorder parameter},
\be
\label{eq:disorder}
\gamma=\frac{1}{G_0^2}\sigma^2 V_c
k_\xi^3/6\pi^2\, .
\ee
In terms of these quantities, Eq.~(\ref{self1}) takes the form,
\be
\label{self2}
\Sig(z)= \; \gamma \int_0^1 dq \; q^4
\bigg( 2 \; \tilde {\cal G}_{L}(q,z)+ 3 \; \tilde {\cal G}_{T}(q,z) \bigg),
\ee
where,
\be
\tilde {\cal G}_{L,T}(q,z)=
\frac{1}{-\frac{z \rho }{G_0 k^2_\xi}+\frac{q^2 \rho}{G_0} v_{L,T}(z)^2
}.
\ee
By further defining the dimensionless squared frequency, $\tilde z$=$z \rho/G_0 k_\xi^2$, and, correspondingly, $\tilde \lambda$=$\lambda \rho/G_0 k_\xi^2$, we can write,
\be
\label{self3}
\Sig(\tilde z)= \; \gamma \int_0^1 dq \; q^4
\bigg(
2 \; \tilde {\cal G}_{L}(q,\tilde z)+3 \; \tilde {\cal G}_{T}(q,\tilde z)
\bigg),
\ee
with
\ba
\label{DefG}
\tilde {\cal G}_{L}(q,\tilde z)&=&\frac{1}{-\tilde z+q^2 \big[ \tilde K_0+\frac{4}{3} \big( 1-\Sig(\tilde z) \big) \big]}
\\
\tilde {\cal G}_{T}(q,\tilde z)&=&\frac{1}{-\tilde z+q^2 \big[ 1-\Sig(\tilde z) \big]}.\nonumber
\ea
The self-consistent Eqs.~(\ref{self3}) and~(\ref{DefG}), must be solved numerically for the self energy, $\Sig(\tilde z)$, and the Green's functions, $\tilde {\cal G}_{L,T}(q,\tilde z)$. The latter can be used in turn to calculate the dimensionless density of eigenvalues as,
\be
\label{dos}
\tilde\rho(\tilde\lambda)=\frac{1}{3\pi}
\Im{  \Big \{ 
{\cal \tilde G}_L(\tilde z)
+2{\cal \tilde G}_T(\tilde z)
\Big\}
},
\ee
with the local Green's functions,
\be
{\cal \tilde G}_{L,T}(\tilde z)=
3\frac{k_\xi^3}{k_D^3}  
\int_0^{k_D/k_\xi}dq q^2  {\cal \tilde G}_{L,T}(q,\tilde z),
\ee
and the Debye cutoff, $k_D=\sqrt[3]{6\pi^2N/V}$. In what follows, for the sake of simplicity, we set $k_\xi=k_D$, obtaining,
\be
{\cal \tilde G}_{L,T}(\tilde z)=
3
\int_0^1dq \; q^2  {\cal \tilde G}_{L,T}(q,\tilde z).
\ee
The numerical solution of the self-consistent equations provides the density of eigenvalues shown in Fig.~\ref{FigA}, for the indicated values of the parameter $\gamma$ (Eq.~(\ref{eq:disorder})). We chose these values to correspond to the chosen temperatures in the simulation (Fig. \ref{inm1}), according to $\gamma3.5 T$, where $T$ is given in Lennard-Jones (LJ) units.
\subsubsection{The transverse contribution}
\label{subsubsect:transverse contribution}
A simple but robust approximation, at least in the low-frequency region, consists in only considering the transverse contribution to the DOS. In this case, the HET-SCBA equation, Eq.~(\ref{self2}) becomes,
\ba
\label{self4}
\Sig(\tilde z)&=& \; \gamma \int_0^1 dq \; q^4 
 \frac{3}{-\tilde z+q^2 \big[ 1-\Sig(\tilde z) \big]}=
 \\
 \nonumber
 &=& \; \gamma \frac{1}{[1-\Sig(\tilde z) ]}\int_0^1 dq \; 3 q^2  \frac{q^2}{q^2 - \frac{\tilde z}{[1-\Sig(\tilde z) ]}}, 
\ea
which, defining the auxiliary function $f(\tilde z) \doteq \tilde z / [1-\Sig(\tilde z)]$, can be also expressed as,
\ba
\label{self5}
&& \Sig(\tilde z) (1-\Sig(\tilde z))  = \gamma \int_0^1 dq \;  3 q^2   \frac{q^2}{q^2 -f(\tilde z)}=
 \\
 \nonumber
 &=& \; \gamma\left\{ 1+ 3 f(\tilde z) \left [ 1 +f(\tilde z)^{1/2} \frac{1}{2} \ln \Big( \frac{f(\tilde z)^{1/2}-1}{f(\tilde z)^{1/2}+1} \Big)\right ] \right\}.
\ea
It is now convenient to write Eq.~(\ref{self5}) as,
\be
\label{self6}
\Sig(\tilde z) (1-\Sig(\tilde z))  = \gamma + 	\tilde z {\cal{F}} (\tilde z),
\ee
where we have defined,
\ba
\label{defgg}
{\cal{F}} (\tilde z) 
&\doteq&\frac{3 \gamma}{1-\Sig(\tilde z)} \left[1 +f(\tilde z)^{1/2}\frac{1}{2}\ln\Big( \frac{f(\tilde z)^{1/2}-1}{f(\tilde z)^{1/2}+1} \Big) \right]=
\nonumber\\
&=&
\gamma \; {\cal \tilde G}_T(\tilde z).
\ea
As the function ${\cal \tilde G}_T(\tilde z)$ reaches a finite value for $\vert \tilde z \vert \rightarrow 0$, Eq.~(\ref{self6}) makes clear that the low-frequency limit of the self-energy is also finite. We may now solve the quadratic Eq.~(\ref{self6}), and obtain~\footnote{We take the solution with the plus sign which, if one goes into the glass with $\gamma$ very small, recovers the Born approximation result \cite{ganter10}.},
\be\label{sigma1}
\Sig(\tilde z)=\frac{1}{2}+\sqrt{\gamma_c-\gamma-\tilde z {\cal{F}} (\tilde z)
},
\ee
with $\gamma_c=\frac{1}{4}$.
The unstable (i.e., liquid) situation is characterized by $\gamma>\gamma_c$. In this case it is convenient re-write Eq.~(\ref{sigma1}) as,
\be\label{sigma2}
\Sig(\tilde z)=\frac{1}{2}+ i\sqrt{\gamma-\gamma_c+\tilde z {\cal{F}} (\tilde z)}.
\ee
The self consistent equations for the transverse contribution to the level density are, therefore, Eqs.~(\ref{defgg}) and (\ref{sigma2}).
\subsubsection{The low-frequency limit}
\label{subsubsect:low-frequency limit}
It is worth noting that the self energy $\Sig(\tilde z)$ satisfies the following properties for vanishing argument,
\ba
\label{prop1}
\Sigma_0&=&\frac{1}{2}+i \sqrt{\gamma-\gamma_c}\nonumber
\\
\Sigma_0^* &=& 1-\Sigma_0
\\
\vert \Sigma_0 \vert^2 &=&\Sigma_0\big(1-\Sigma_0\big)= \gamma,
\nonumber
\ea
where we have denoted, $\Sigma_0 \doteq \Sig(0)$). By defining $\xi=\xi'+i\xi''=\Sigma_0^{1/2}$, we next obtain,
\ba
\label{prop2}
\xi'&=&\frac{1}{2}\sqrt{\sqrt{1+4(\gamma-\gamma_c)}+1}
\\
\nonumber
\xi''&=&\frac{1}{2}\sqrt{\sqrt{1+4(\gamma-\gamma_c)}-1},
\ea
and by combining both equations,
\be
\xi''=\frac{1}{\xi'}\Sigma_0''=\frac{1}{\xi'}\sqrt{\gamma-\gamma_c}\,.
\ee

To finalize the list of properties of the self energy at small frequency, we observe that, besides the constant $\Sigma_0$, the leading term is linear, as one can conclude from the relation,
\be
\label{prop3}
\Sig(\tilde z) = \Sigma_0 \Big( 1+ i \frac{3}{2} \frac{1}{\sqrt{\gamma-\gamma_c}} \tilde z  \Big) + \mathcal{O}(\tilde z^{3/2}).
\ee
The density of eigenvalues $\lambda$, considering the transverse contributions only, is therefore given by,
\be
\label{dos2}
\tilde\rho(\tilde\lambda) 
= \frac{2}{3 \pi} \Im{  \Big \{
	{\cal G}_{T}(\tilde z) \Big\} }
= \frac{2}{3 \pi\gamma} \Im{  \Big \{
	{\cal F}_{T}(\tilde z) \Big\} }.
\ee
We finally analyze the imaginary part of ${\cal{F}} (\lambda)$ in Eq.~(\ref{defgg}) at low frequency. For small $\lambda$, the argument of the logarithm becomes -1, therefore its real part vanishes while the imaginary part become constant and equal to $\pi$,
\ba
\label{dos5}
{\cal{F}} ''(\lambda) &\approx& \Im{  \Big \{   \frac{3 \gamma}{[1-\Sig(\lambda)]} \Big( 1 + i \frac{\pi }{2} f(\lambda)^{1/2} \Big) \Big\} }=
\\
\nonumber
&=& \Im{  \Big \{   \frac{3\gamma}{[1-\Sig(\lambda)]} \Big( 1 +  i \frac{\pi }{2} \frac{\lambda^{1/2}}{[1-\Sig(\lambda)]^{1/2}} \Big) \Big \}}.
\ea
\subsubsection{The zero-frequency value and the cusp}
\label{subsubsect:zero-frequency value and the cusp}
It is now straightforward to obtain the density of eigenvalues at $\lambda=0$ as,
\be
\tilde\rho_o \doteq
\tilde\rho(\tilde\lambda\!=\!0)
=\frac{2}{\pi}\Im\bigg\{\frac{1}{1-\Sigma_o}\bigg\}
=\frac{1}{\pi}\frac{\sqrt{4\gamma-1}}{\gamma}.
\label{rhozero}
\ee
Interestingly, this value is not approached linearly by our theory, but rather in a non-analytic way, producing a characteristic {\it cusp} already observed in~\cite{taraskin02a}.
%
%
\begin{figure}[t]
\centering
\includegraphics[width=0.49\textwidth]{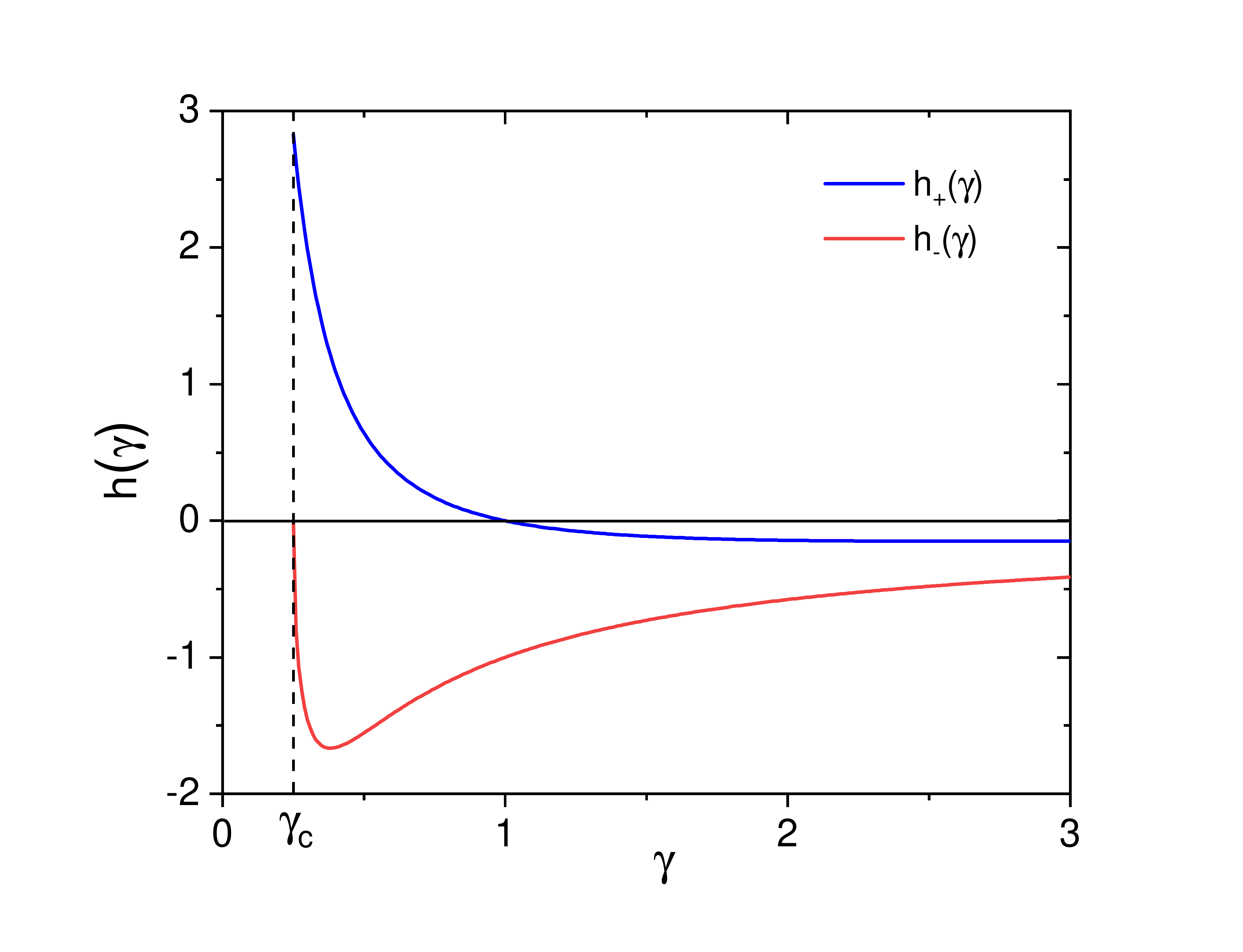}
\caption{Prefactor of the $\lambda^{1/2}$ term in the density of level $\rho(\lambda)$ (Eq.~(\ref{ll})) for positive (blue line) and negative (red line) values of $\lambda$. Only the contributions of the transverse modes to the level distribution are considered here. The prefactor is always negative (thus the cusp is upward) in the negative $\lambda$ region. For $\lambda>0$ the direction of the cusp depends on the value of $\gamma$. At $\gamma$ close to $\gamma_c$(=1/4), corresponding to low temperatures, the cusp is downward, on increasing $\gamma$ (in our approximation, at $\gamma$=1) the cusp changes direction.
\label{sharp}
}
\label{f0}
\end{figure}
Indeed, beside the constant value reached at $\tilde\lambda$=0, the leading term of the function ${\cal{F}} (\tilde\lambda)$ is proportional to $\tilde\lambda^{1/2}$. In the low frequency expansion of Eq.~(\ref{dos5}) we can therefore safely substitute $\Sig(\tilde\lambda)$ with $\Sigma_0$, obtaining,
\ba
\label{dos5b}
{\cal{F}} ''(\lambda) &\approx& \Im{  \Big \{   \frac{3\gamma}{[1-\Sigma_0]} \Big( 1 +  i \frac{\pi }{2} \frac{\tilde\lambda^{1/2}}{[1-\Sigma_0]^{1/2}} \Big) \Big \} }
\\
\nonumber
&=& \Im{  \Big \{   3 \Sigma_0 \Big( 1 + i \frac{\pi }{2} {\Sigma_0^{1/2}} \frac{1}{\sqrt{\gamma}}{\tilde\lambda^{1/2}}\Big) \Big \} }
\\
\nonumber
&=& \Im{  \Big \{   3 (\Sigma'_0+i\Sigma''_0) \Big( 1 + i \frac{\pi }{2} (\xi' + i\xi'') \frac{1}{\sqrt{\gamma}}{\tilde\lambda^{1/2}}\Big) \Big \} }.
\ea
We see that, for $\tilde\lambda$ small but $>0$,
\be
\label{lp}
{\cal{F}} ''(\tilde\lambda)= 3 \Sigma''_0 + \tilde\lambda^{1/2} \frac{3 \pi}{2 \sqrt{\gamma}}\Big\{ \xi' \Sigma_0' - \xi'' \Sigma_0'' \Big\},
\ee
while, for negative $\tilde\lambda$ we find,
\be
\label{lm}
{\cal{F}} ''(\tilde\lambda)= 3 \Sigma''_0 + \vert \tilde\lambda \vert^{1/2} \frac{3 \pi}{2 \sqrt{\gamma}}\Big\{ -\xi' \Sigma_0'' - \xi'' \Sigma_0' \Big\}.
\ee
Summing up, for $\gamma_c$=1/4 we can write,
\be
\label{ll}
\tilde\rho(\tilde\lambda)=\frac{2}{3\pi\gamma}
{\cal{F}} ''(\tilde\lambda)=\tilde \rho_o + h_\pm(\gamma) \; \vert \tilde \lambda \vert^{1/2},  
\ee
where the function $h_+(\gamma)$ ($h_-(\gamma)$) must be used for the positive (negative) $\lambda$-regions, and are defined as,
\be
\label{h} 
h_\pm(\gamma) = \frac{\gamma^{-3/2}}{4}\left[- \sqrt{(4\gamma-1)(2 \gamma^{1/2} \mp 1)} \pm\sqrt{(2 \gamma^{1/2} \pm 1)}  \right].
\ee
A non-analytical behavior of the density of level is therefore clear in Eq.~(\ref{ll}), with an asymmetric square-root cusp centered at $\lambda=0$. The functions $h_\pm(\gamma)$, determining both direction and sharpness of the cusp, are shown in Fig.~\ref{sharp}. We see that for negative values of $\lambda$, we always obtain a upward $\lambda^{1/2}$ cusp. For $\lambda>0$, in contrast, we find a positive rise proportional to $\lambda^{1/2}$ in the region where $\gamma$ is small (i.e., at low temperatures), whereas for large $\gamma$ an upwards $\lambda^{1/2}$ cusp is obtained. The transition between these two regimes occurs at $\gamma=1$.

\begin{figure}[t]
\centering
\includegraphics[width=0.49\textwidth]{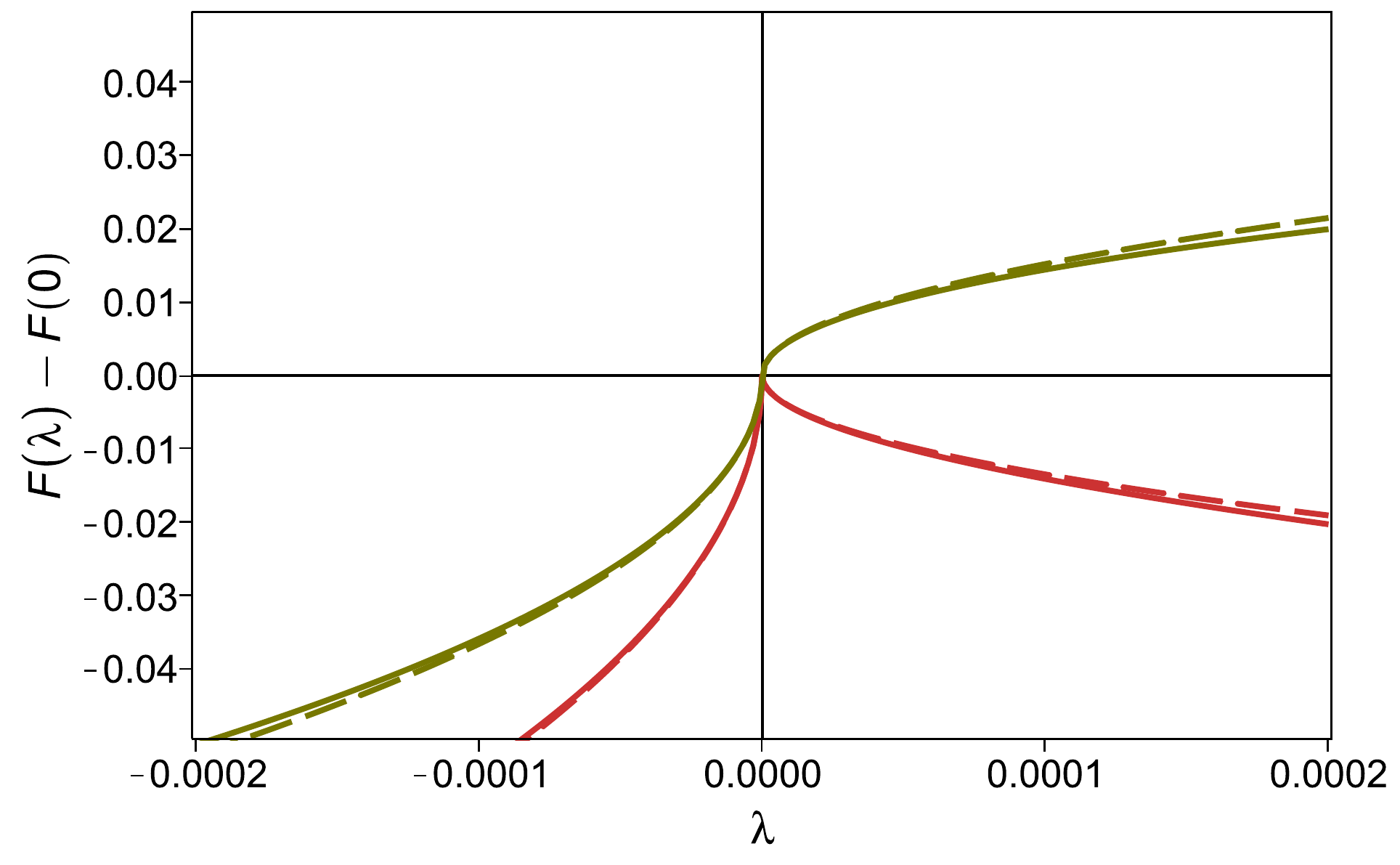}
\caption{Test of the low frequency approximation for the function ${\cal{F}} ''(\lambda) $. The dashed lines are approximate solution from Eqs.~(\ref{ll}) and~(\ref{h}), while the full lines are obtained by a numerical solution of the self-consistent Eqs.~(\ref{sigma2}) and~(\ref{defgg}). The red lines refer to $\gamma$=0.5, the green to $\gamma$=2.}
\label{test}
\end{figure}

The validity of the low-frequency approximation of Eqs.~(\ref{ll}) and~(\ref{h}) is demonstrated in Fig.~\ref{test}, where the "exact" solution of the self-consistent Eqs.~(\ref{sigma2}) and~(\ref{defgg})(full line) is compared with the approximated one (dashed line), for the two indicated values of $\gamma$. We conclude that the description of the low frequency level density with a cusp is valid and robust. 

The origin of  the non-analytic singular behavior discussed above can be traced back to the sum rule $\sum_jH_{ij}^{\alpha\beta}=0$ for the Hessian~\cite{taraskin02a}, due to the global translation invariance. In our elastic model the sum rule is transformed to the double spatial derivative in the elastic wave Eq.~(\ref{eqmo1}), or in the Helmholtz Eqs.~(\ref{helm2}), again reflecting global translation invariance.  Technically,  the $\lambda^{1/2}$ singularity  stems from vanishing values of the wave-number $k$ in the integrand of the SCBA Eq.~(\ref{self1}), corresponding to density and stress fluctuations of very large extent. We convinced ourselves that the singularity at $\lambda=0$ would disappear if an infrared cutoff $k_0$ is introduced at the lower limit of the integral in {Eq.~(\ref{self1})}, which mimics the effect of a finite sample size. As a consequence, in numerical simulations at finite system sizes, one expects a finite slope at $\lambda=0$, at variance with the $\lambda=0$ singularity of the SCBA. We will address this point below.

Concluding the presentation of the theoretical results, and specifically of the existence of a cusp in the vibrational levels spectrum,  we emphasize that this result is an additional evidence of the intrinsic weakness of exceedingly crude approximations sometimes employed in the literature, including the prescription $\rho(\lambda)\sim\textit{const}$ (or, equivalently, $g(\omega) \sim \vert \omega \vert$)~\cite{zaccone21,dehong22}. Indeed, in considering this approximation many details of the liquid dynamics are plainly ignored.
\section{Numerical methods}
\label{sect:numerical methods}
In the following Sections we will compare the above theoretical predictions to the outcome of extensive computer simulations. Here, we give details about the atomic liquid model used, the molecular-dynamics (MD) simulations performed, and the numerical tools employed in the INM analysis. Below, all observables are expressed in Lennard Jones units by setting the fundamental quantities, $m$, $\sigma$, $\epsilon$ (mass, length, and energy, respectively), and the Boltzmann constant $k_B$ to 1, while $\sigma\sqrt{m/\epsilon}$ is the unit of frequency. 

We have considered systems formed by equimolar binary mixtures of $N$ soft-spheres of type $A$ (small) and $B$ (large), respectively, with $N_A=N_B=N/2$. The spheres interact via the pair-wise potential,
\begin{equation}
V(r)/\epsilon=\sigma^{12}\left(\frac{1}{r^{12}}-\frac{1}{r_c^{12}}\right)\times\left[ 1-\left(\frac{r}{r_c}\right)^2\right]^2,
\label{eq:potential}
\end{equation}
which is the product of the usual soft sphere potential ($\propto r^{-12}$), cutoff and shifted at $r_c=1.5$, and a tapering function that ensures continuity of both potential and forces at the cutoff. Here $r=r_{ij}$ is the distance between the particles $i$ and $j$, $\epsilon_{AA}=\epsilon_{BB}=\epsilon_{AB}=1$, $\sigma_{AA}=0.90$, $\sigma_{BB}=1.09\simeq 1.2\, \sigma_{AA}$, and $\sigma_{AB}=(\sigma_{AA}+\sigma_{BB})/2\simeq 1$. Masses are $m_A=m_B=m=1$. All simulations have been conducted by using the high-performance-computing simulation code LAMMPS~\cite{thompson2022lammps}. 

We have considered 7 system sizes, $N=10^3\times 2^n$ with $n=0,\ldots, 7$ ($n=6$ excluded), ranging from the situation where the entire set of the Hessian eigenvalues can be calculated comfortably (small $N$), to that where we are only able to obtain the vibrational density of state in a limited frequency range around $\lambda\simeq 0$ (large $N$), although with high statistical accuracy. We have fixed the simulation box sizes, $L$, by matching the number density $\hat{\rho}=N/L^3=1$, for all systems. We considered 24 values of temperature, $T$, in the range $[0.08:1]$. Starting at $T=1$, the system was slowly annealed to $T=0.08$ in steps of $\delta T=0.04$. At each temperature, a total of $12\times 10^6$ time steps were performed, consisting of a $2\times 10^6$ thermalization period, followed by the production run with the dumping of the system configurations used in the subsequent analysis. The runs were carried out in the $(NVE)$ ensemble with a numerical integration time step $\delta t=0.002$. 

We have implemented the calculation of the Hessian matrix, Eq.~(\ref{in1}), in JAX~\cite{jax2018github}, which can automatically differentiate native Python and NumPy functions. $H^{\alpha\beta}_{ij}$ has been evaluated at each value of $T$ and $N$ on a number of independent istantaneous configurations sufficient to reach the desired statistical accuracy, and subsequently diagonalized to extract the eigenvalues. We have employed the standard scipy.linalg~\cite{2020SciPy-NMeth} linear algebra functions at small values of $N$, while for $N\ge 16000$ we used RALEIGH~\cite{ovtchinnikov2008computing}, a Python implementation of the block Jacobi-conjugated gradients algorithm for computing eigenvalues and eigenvectors of large-scale real symmetric and Hermitian problems in selected eigenvalue ranges. We have eventually estimated the INM spectra, $\rho(\lambda)$, by constructing histograms of the obtained eigenvalues. 
\begin{figure}[t]
\centering
\includegraphics[width=0.49\textwidth]{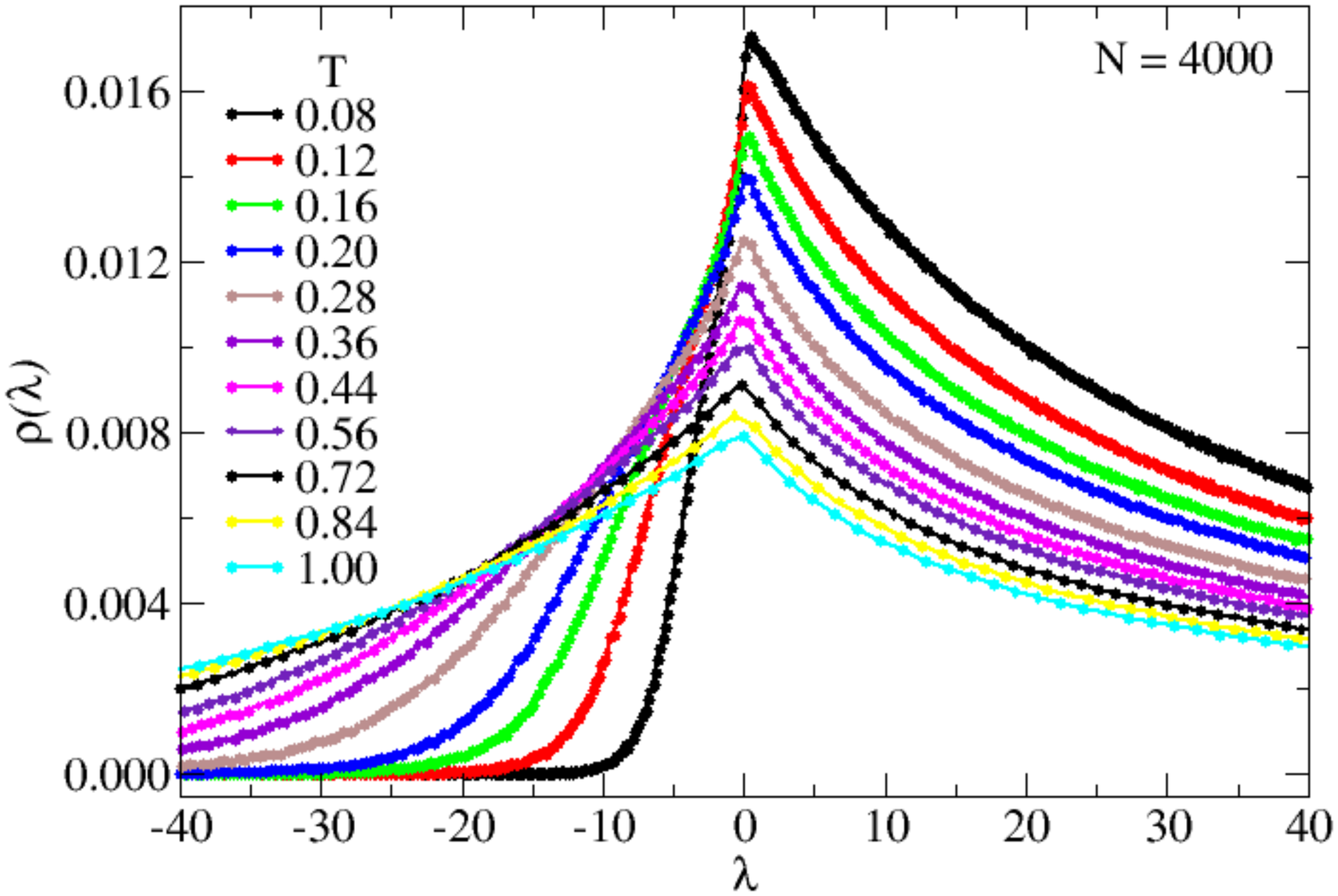}
\caption{Simulated INM spectra, $\rho(\lambda)$ , for $N=4\times 10^3$ at the indicated values of $T$. For the investigated soft sphere system, $T_{MCT}\simeq 0.05$. All quantities are expressed in LJ units. These data are the numerical counterpart of those shown in Fig.~\ref{FigA}. The number of unstable modes decreases significantly upon cooling and, as a consequence, distributions which are quite asymmetric around $\lambda=0$ at high $T$, becomes consistently increasingly asymmetric by decreasing $T$, as expected. These data are discussed at length in the main text.}
\label{inm1}
\end{figure}
\begin{figure}[t]
\centering
\includegraphics[width=0.49\textwidth]{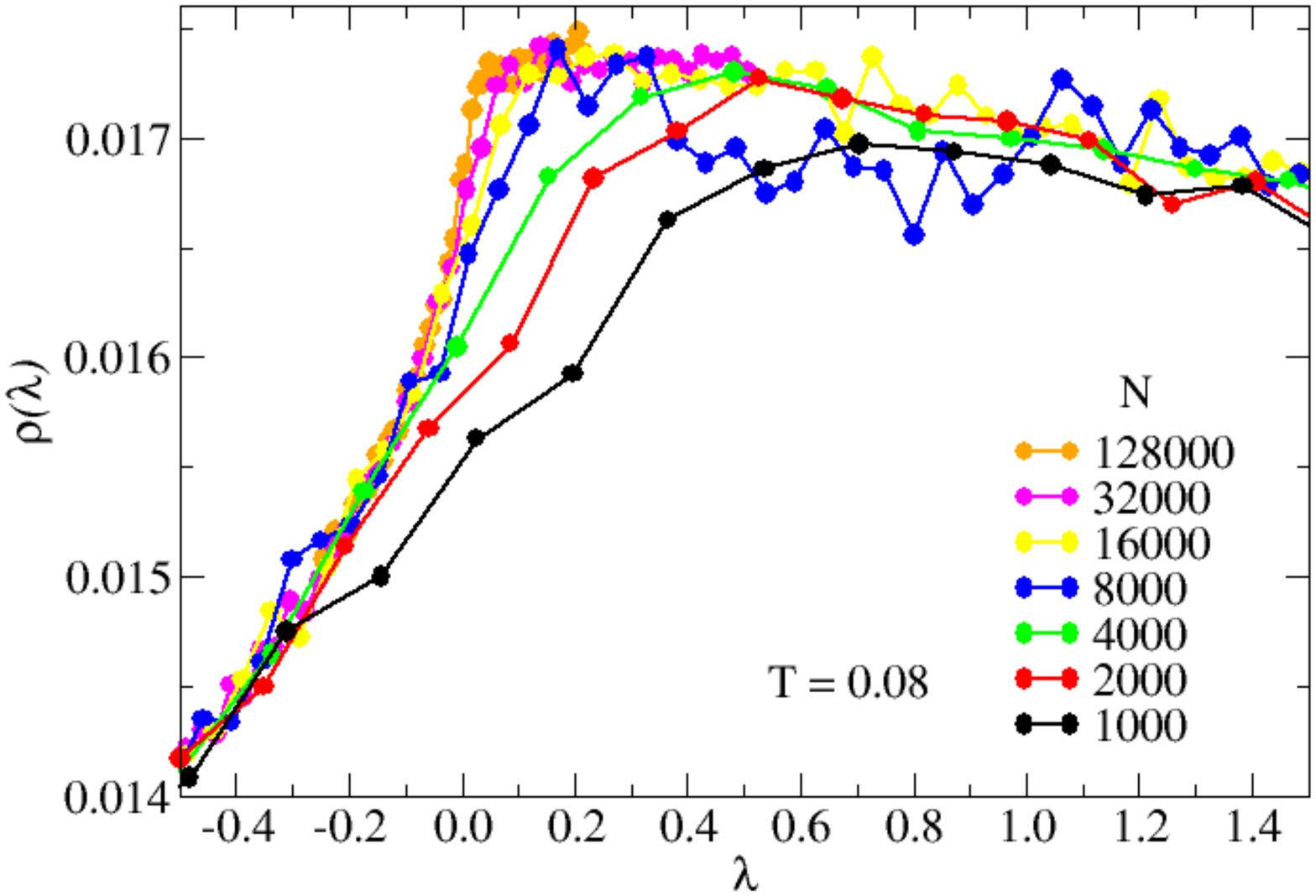}
\caption{Effect of the finite system size on the low-$\lambda$ INM spectral shape. Simulated INM spectra, $\rho(\lambda)$, at the lowest investigated temperature, $T=0.08$, for the indicated values of $N=10^3\times 2^n$, with $n=0,\ldots, 7$ ($n=6$ excluded). At the smaller values of $N$ the $\rho(\lambda)$ intensities increase continuously on the unstable side, crossing $\lambda=0$ and eventually reaching maxima at positive values of $\lambda$. The positions of the maxima, however, continuously shift toward $\lambda=0$ on increasing $N$. At the highest investigated values of $N$ the cusp-like non-analytical behavior starts to develop, as discussed into details in the main text.   
}
\label{f2}
\end{figure}
\section{Numerical Results}
\label{sect:numerical results}
We now present a comparison of the simulated spectra of the soft-sphere liquid with the predictions obtained by the unstable version of HET-SCBA theory, together with an in-depth discussion of the data especially in light of the discussion of Sect.~\ref{subsubsect:zero-frequency value and the cusp}. In particular, we focus on the finite-size effects on the distribution of the eigenvalues of the Hessian matrix, and on the details of the shape of $\rho(\lambda)$ in different conditions, both in the limit $\lambda\rightarrow 0$ and for intermediate values of $\lambda$.
\subsubsection{Overall INM spectrum}
\label{subsubsect:overall inm spectrum}
We show in Fig.~\ref{inm1} the calculated INM spectra of the investigated soft-spheres liquids, for the indicated values of $T$, for $N=$~4000. The lowest considered temperatures are just above the arrest temperature (we estimate, $T_{MCT}\simeq$ 0.05), while the highest values of $T$ encompass the high-fluidity regime, up to $T\simeq 20\;T_{MCT}$. We observe that the spectrum changes from an almost stable situation with a very small fraction of negative eigenvalues at low $T$ (black symbols), to a distribution which is almost symmetric with respect to positive and negative $\lambda$'s. These $T$-dependent data are very similar to those predicted by the HET-SCBA shown in Fig.~\ref{FigA} for the indicated values of the disorder parameter, $\gamma$. As said above, we chose these values according to $\gamma=3.5 T$. While a more detailed comparison of the numerical and analytical $\rho(\lambda)$ curves would give a slight deviation from this linear law, the overall trend is given correctly. We already mentioned that in our earlier simulation \cite{mizuno2019impact}, in which the temperature dependence of the variance $\sigma^3\propto \gamma$ was evaluated, an increase with temperature was observed at the highest considered temperatures.

We turn now to a comparison between the simulated $\rho(\lambda)$ curves (Fig. \ref{inm1}), and those predicted by HET-SCBA (Fig.~\ref{FigA}). The only notable difference between the two sets of data is that in the spectra of Fig.~\ref{FigA} the negative-$\lambda$ tails are truncated, a typical effect of the SCBA approximation, which provides spectra extending on a support of finite width. No cutoff is visible in the simulation data, as expected.

HET-SCBA therefore seems to satisfactorily reproduce the numerical INM spectra modifications with $T$, including the transition from a stable situation at $T\simeq T_{MCT}$ to a quite symmetric distribution at high $T$, while, as expected due to the normalization, the peaks heights decrease with $T$. We also note that the maximum of the distribution is always found at $\lambda_M>0$, in contrast to the intuitive expectation $\lambda_M=0$. We will better discuss this point below. 
\begin{figure}[t]
\centering
\includegraphics[width=0.4\textwidth]{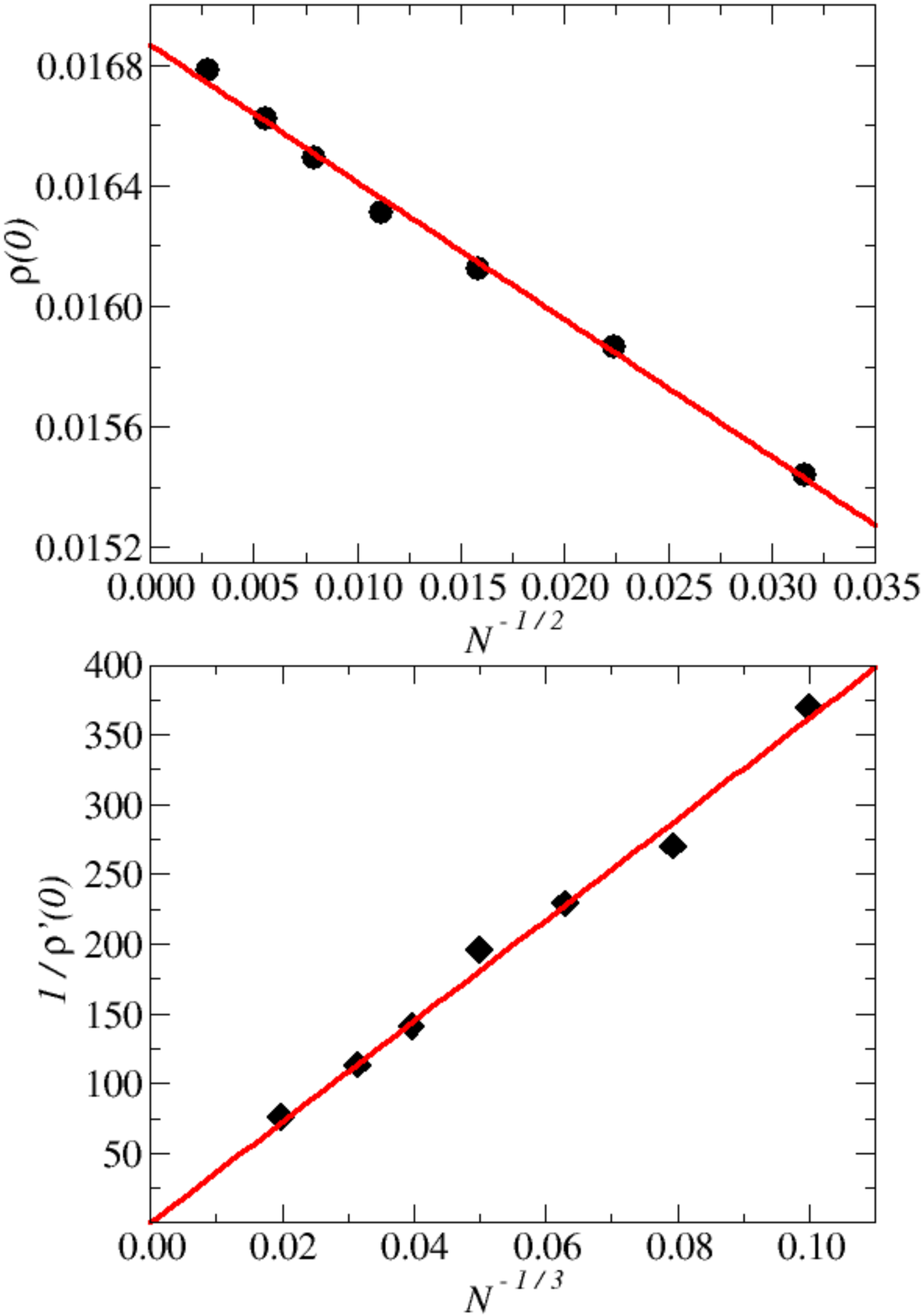}
\caption{Finite-size scaling of values and derivatives of the INM spectra at $\lambda=0$. The data have been determined by fitting linearly the data of Fig.~\ref{f2} around $\lambda=0$. {\it Top:} Values of the INM spectra $\rho(\lambda=0)$ at $T=0.08$, as a function of $N^{-1/2}$. Data are indicated by the symbols, the solid line is a linear guide for the eyes. {\it Bottom:} Inverse of the derivative of the INM spectra at $\lambda=0$, represented as a function of $N^{-1/3}$. The details of the observed scaling with the system size are discussed in the main text. 
}
\label{f3}
\end{figure}
\subsubsection{Finite-size effects}
\label{subsubsect:finite size effects}
We now quantify the effect due to the finite size of the simulated systems on the $\rho(\lambda)$. In Fig.~\ref{f2} we show the details of $\rho(\lambda)$ in the very low-$|\lambda|$ region, for the indicated system sizes $N$, at the lowest investigated temperature, $T=0.08$ (corresponding to the black symbols in Fig.~\ref{inm1}). We observe a strong $N$-dependence of the spectral shape, with a slope for $\lambda<0$ which consistently increases by increasing $N$. Importantly, the theory predicts a derivative which diverges at $\lambda$=0 (see, for instance, Eq.~(\ref{ll})), which cannot be directly observed in our simulation. Our data, however, point in the correct direction, allowing us to qualitatively conclude that our simulation is consistent with the theory in the thermodynamic limit, $N\rightarrow\infty$. 

To better emphasize this important point, we have extracted the main features of $\rho(\lambda)$ close to $\lambda=0$ by a linear fit of the data of Fig.~\ref{f2} restricted to the negative (unstable) $\lambda$-branch, and up to the position of the distribution maximum, always localized at positive values of $\lambda$, as already noted above. We show in Fig.~\ref{f3} (top) the obtained values for $\rho(\lambda=0)$, represented as a function of $N^{-1/2}$, and for the inverse derivative $1/\rho'(0)$ shown as a function of $N^{-1/3}$ (bottom). 

In the $N \rightarrow \infty$ limit, we find that $\rho(0)$ extrapolates to a finite value $\simeq 1.686\times 10^{-2}$ (see Eq.~(\ref{rhozero})), while $\rho'(0)$ diverges, as expected from the presence of the cusp, Eq.~(\ref{ll}), for $\gamma$ close to criticality.  HET-SCBA therefore truly succeeds in identifying quantitatively the zero-energy singularity in the simulated energy spectrum. We now demonstrate that it also provides quantitative information about the cusp-like character of the singularity. Here, the possibility of analyzing quite large system sizes is crucial. 
\begin{figure}[t]
\centering
\includegraphics[width=0.49\textwidth]{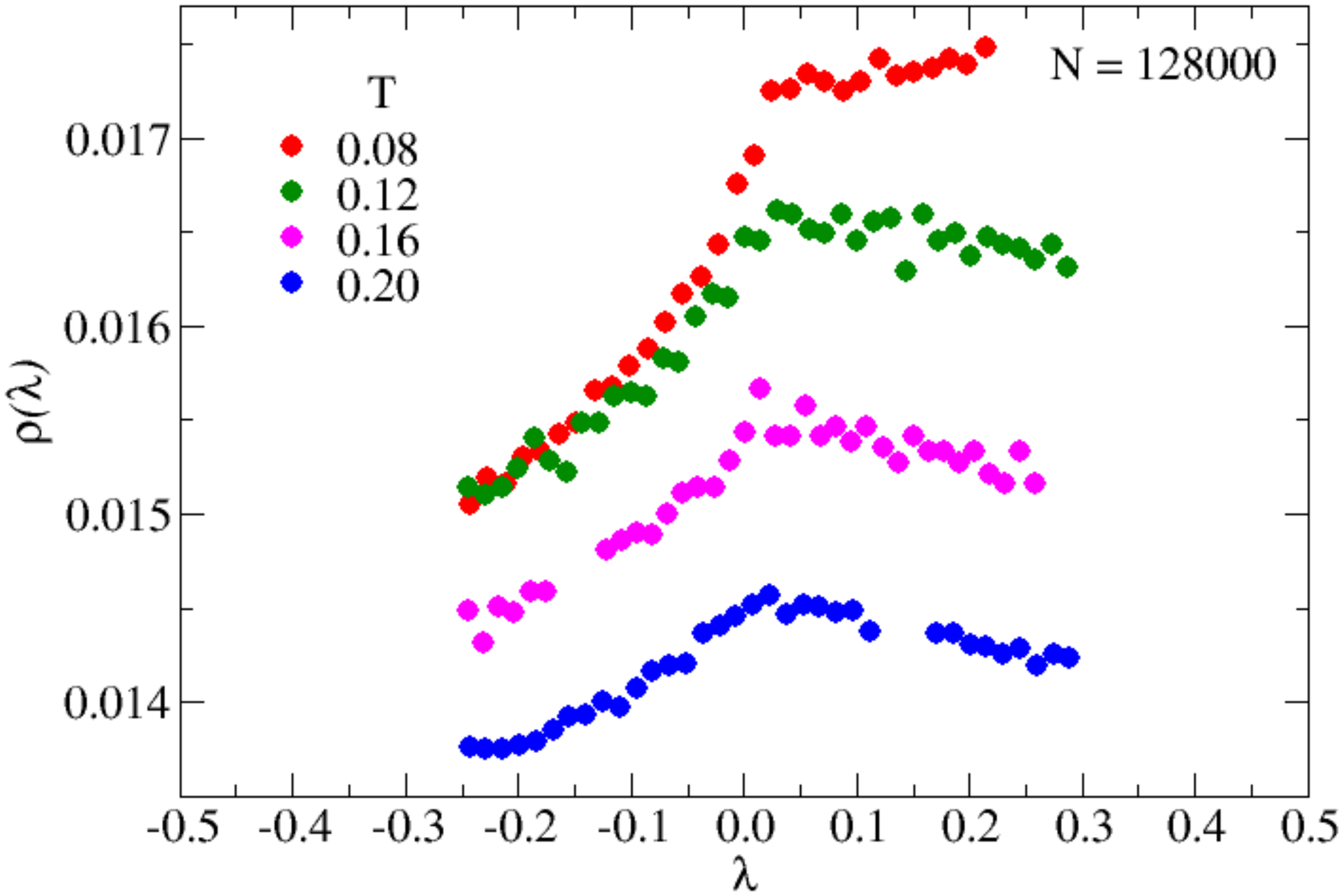}
\caption{Details of the low frequency region of the INM spectra. We plot $\rho(\lambda)$ for N=128000, at the four lowest values of $T$. All details of the data, both in the unstable and stable $\lambda$-regions, are discussed in the main text.}
\label{fig7}
\end{figure}
\subsubsection{The cusp at $\lambda$=0}
\label{subsubsect:the cusp}
We are now in the position to analyze quantitatively the shape of the simulated $\rho(\lambda)$ curves at small values of $|\lambda|$ on both the stable and unstable regions, and its dependence on temperature or, equivalently, on the disorder parameter, $\gamma$. We show in Fig.~\ref{fig7} the details of the $\rho(\lambda)$ at small $|\lambda|$ values  for the largest investigated system size $N$=128000, at the indicated low temperatures. 

\begin{figure}[t]
\centering
\includegraphics[width=0.49\textwidth]{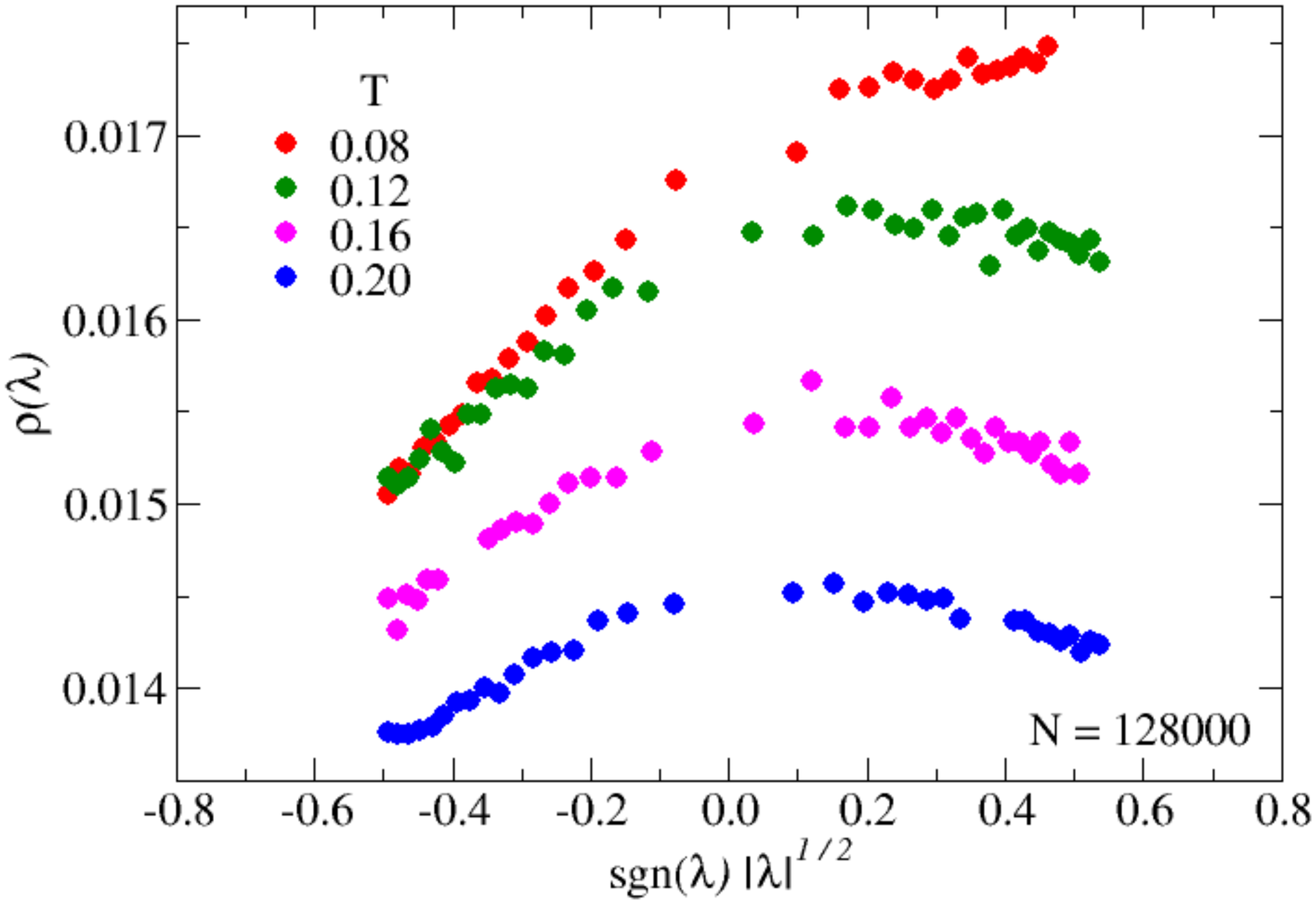}
\caption{Same data of Fig.~\ref{fig7}, now plotted as a function of $\text{sgn}(\lambda)|\lambda|^{1/2}$. The consequences of this representation are discussed in details in the main text.}
\label{fig8}
\end{figure}

With reference to Eqs.~(\ref{ll}) and~(\ref{h}), at the lowest $T=0.08$ (red symbols, corresponding to $\gamma\approx\gamma_c$), we find that $h_+(\gamma)>0$, while $h_-(\gamma)$, in contrast, is negative in all cases, as predicted by Eq.~(\ref{h}). Also, on increasing $T$ ($\gamma$), the sign of $h_+(\gamma)$ reverses, as expected, indicating that  already at $T$=0.12 we  have $\gamma >1$. As illustrated in Fig.~\ref{fig8}, by plotting $\rho(\lambda)$ against $\mathrm{sign}(\lambda) \vert \lambda \vert^{1/2}$, the curves of Fig.~\ref{fig7} are rectified (according to Eq.~(\ref{ll})), and the slopes for $\lambda$ positive and negative are different. This difference tends to vanish on increasing $T$ ($\gamma$), where the distribution becomes smoother (see blue points in Fig.~\ref {fig8}). 

We conclude that the simulations performed on a soft sphere model confirm both qualitatively and quantitatively the prediction of the INM made on the basis of the  HET extended to the liquid case. In particular, we find that the critical behavior of the distributions $\rho(\lambda)$ in the vicinity of the zero-energy spectral singularity is very satisfactorily described by the theory. HET-SCBA seems indeed to adequately grasp the behavior of a realistic model of an atomic fluid.
\begin{figure}[t]
\centering
\includegraphics[width=0.49\textwidth]{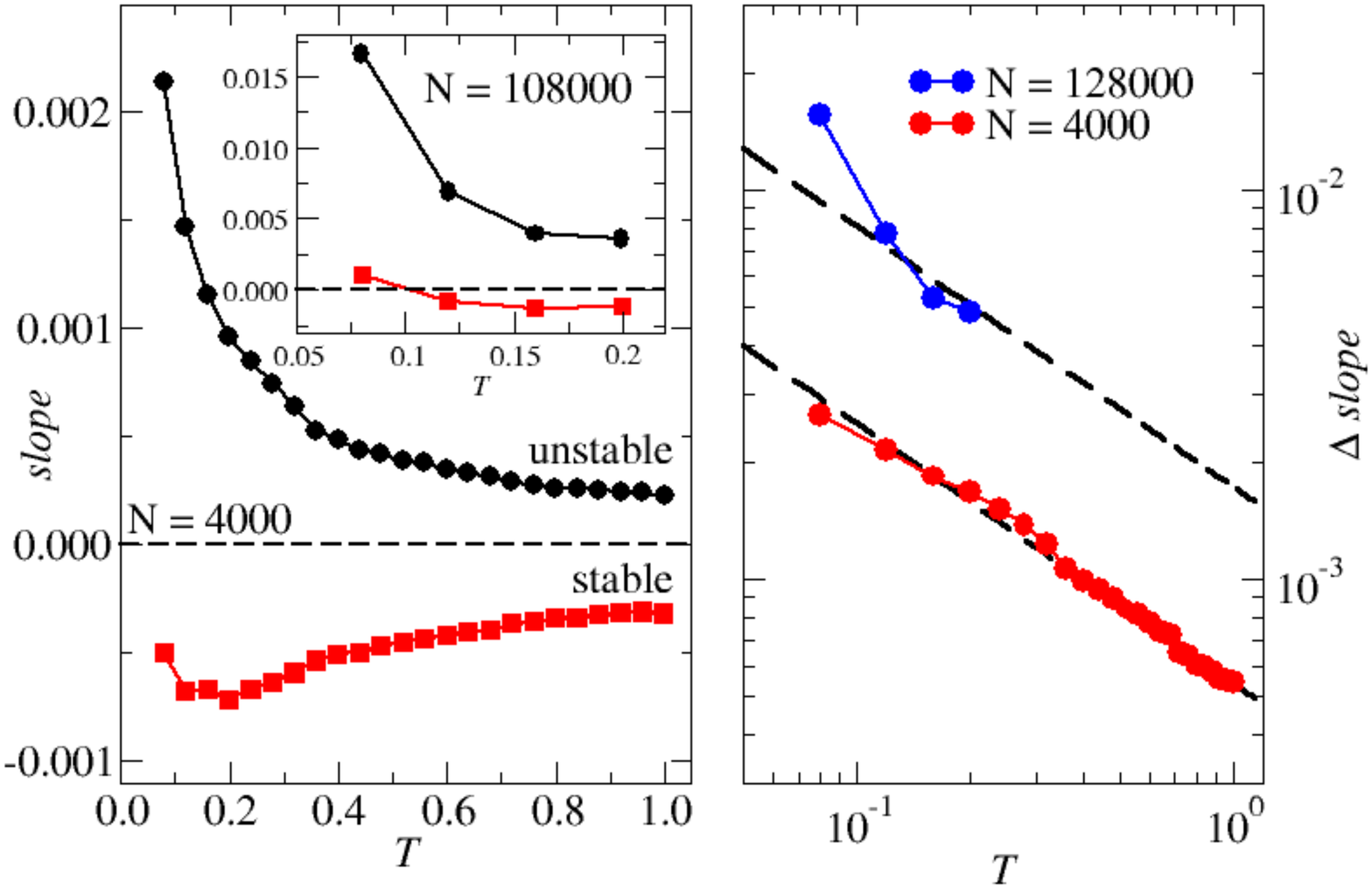}
\caption{Local slopes of $\rho(\lambda)$ close to $\lambda=0$, both in the stable and unstable regions. {\it Left, Main panel:} Local slope for $N=4000$ as a function of temperature, $T$. The slopes on the stable (squares) and unstable (circles) regions around $\lambda=0$ have been determined by a simultaneous non-linear fit of the two regions, as discussed in the text. In the unstable region the slope is always positive, and consistently increases upon cooling. In the stable region, in contrast, the slope is always negative, increasing (in absolute value) on lowering $T$, going through a maximum at $T\simeq 0.2$ and eventually decreasing.) {Left, Inset:} Same that in the main panel, for $N=128000$. The behavior is similar but the decreased impact of finite-size effects indicates that  the slope of the stable region becomes positive at $T\simeq$~0.1. {\it Right:} Slopes differences $\Delta S= S_{\text{s}}-S_{\text{u}}$ for both values of $N$, together with the $\propto T^{-2/3}$ dependence (dashed lines). All data are discussed at length in the main text.}
\label{exponent}
\end{figure}
\subsubsection{The slope of $\rho(\lambda$) at intermediate frequency}
\label{subsubsect:intermediate frequency}
An interesting detail of the spectra obtained in numerical simulations, which cannot be easily extracted from analytical calculations, are the strength and sign of the slope of the $\rho(\lambda)$ curves at small but finite $|\lambda|$ (thus away from the cusp region), in both the stable and unstable regions. We have estimated both slopes by a single non-linear fit to an empirical function of $\lambda$, around $\lambda=$~0~\footnote{\textcolor{black}{We have used the model $\rho(\lambda)=[e^{-\alpha\lambda}(c_0+c_1\lambda)+e^{\alpha\lambda}(c_2+c_3\lambda)] / (e^{-\alpha\lambda}+e^{\alpha\lambda})$. The values of the slopes are directly provided by $c_1$ and $c_3$. We have checked that fitting procedures based on different model functions, including linear models, provide very similar results.}}. 

We plot in the left panel of Fig.~\ref{exponent} the estimated slopes as a function of $T$ for $N=$~4000 (main panel) and 128000 (inset). In the unstable region (circles) the slope is always positive, and consistently decreases upon heating. In the stable region, in contrast, the slope is negative. Interestingly, it increases (in absolute value) on cooling, goes through a maximum at $T\simeq$~0.2 and eventually decreases approaching $T_{MCT}$. (Very similar results have been reported in~\cite{schirm22}). The data for large $N$ in the inset provide substantially the same picture, with the additional information that the slope of the stable region becomes positive at $T\simeq$~0.1, as already noted above.

Unfortunately we do not have an intuition of the mechanism behind the observed behavior of the stable region slope, which we postpone to future work. As more work is needed to extract from the theory a clear explanation of the data shown in the right panel of Fig.~\ref{exponent}. Here, we plot the difference $\Delta S = S_{s} - S_{u}$ of the stable and unstable slopes, on a double-logarithmic scale as a function of $T$ for both sets of data. As already observed by some of us~\cite{schirm22}, the data follow a $T^{-2/3}$ law, in the entire investigated temperature range. A satisfactorily explanation of this observation is lacking.
\section{Conclusions}
\label{sect:conclusions}
Melting of a crystal is accompanied by the suppression of sharp features of the vibrational spectrum. For instance, the Van Hove singularities in the density of states, $g(\omega)$, (associated to the critical points of the dispersion, $\omega(q)$) disappear, and are substituted by the appearance of a broad distribution of vibrational modes, including both stable {\it and} unstable excitations. Interestingly, the reverse process can be very different. This is the case for systems that can be supercooled, i.e., where it is possible to avoid crystallization by a succession of equilibrium liquid states extending below the melting temperature. In this process, the high fraction of unstable modes contained in the spectrum of the Hessian matrix of the liquid vanishes at $T_{MCT}$, which is associated to a dynamical phase transition, and marks the boundary where a saddle-dominated PEL liquid-like kinetics gives way to a minima-dominated flow, eventually leading to glass formation. The limiting shape of the vibrational spectrum approaching $T_{MCT}$ turns out to be surprising, with the appearance of a zero-energy spectral singularity, now unrelated to the $\omega(q)$, with a cusp-like character, and an origin which can be traced back to the most obvious of the symmetries, translational invariance. 

Despite the generality of this phenomenon also discovered, for instance, in Anderson hamiltonians with off-diagonal disorder \cite{eilmes98}, where a similar cusp in the electronic DOS at $E=0$ is observed, an analytic description
of this phenomenon until now has not been provided. Here, we have addressed in depth the characterization of the zero-energy singularity, by putting together insight coming from theoretical advances and extensive computer simulations. We have presented  a generalization to the liquid state of the heterogeneous elastic theory, by integrating in the elastic medium description of the instantaneous normal modes spectrum an explicit temperature dependence of the spatial shear modulus fluctuations. Specifically we have been able not only to demonstrate the existence of a disorder-induced zero-energy spectral singularity, but also fully characterize the disorder dependent spectral shape, in both stable and unstable energy regions. We have next tested the theoretical predictions against extensive Molecular Dynamics simulations of an atomic glass-forming liquid, confirming the relevance of the developed mean-field theory for realistic systems.  

In a nutshell, we have demonstrated that focusing to the original INM spectrum $\rho(\lambda)$ reveals much more interesting information than by transforming it to the $\omega=\sqrt{\lambda}$ spectrum, which completely suppress most relevant zero-energy features of the spectra. Our findings also pose serious constraints on the allowed shape of $\rho(\lambda)$ employed in theoretical developments, sometimes at variance with superficial recent proposals. They finally constitute a very general step forward for all condensed matter systems described by non-negligible off-diagonal disorder.  
\section{acknowledgments}
We are grateful to Jean-Louis Barrat, Matteo Paoluzzi and Richard M. Stratt for helpful discussions and suggestions. S. M. acknowledges support by the project Heatflow (Grant No. ANR-18-CE30-0019-01) funded by the French national funding agency “Agence Nationale de la Recherche”.
\bibliography{glassybib}
\end{document}